\documentclass[acmsmall,screen]{acmart}
\usepackage[utf8]{inputenc}
\usepackage[T1]{fontenc}
\usepackage{graphicx}
\usepackage{grffile}
\usepackage{longtable}
\usepackage{wrapfig}
\usepackage{rotating}
\usepackage[normalem]{ulem}
\usepackage{amsmath}
\usepackage{textcomp}
\usepackage{capt-of}
\usepackage{hyperref}
\usepackage{listings}
\usepackage{listings}
\usepackage{color}
\usepackage[utf8]{inputenc}
\usepackage{syntax}
\usepackage{amsthm}
\usepackage{mathrsfs}
\usepackage{subfigure}
\usepackage{tabularx}
\acmConference[miniKanren 2024]{The Seventh miniKanren and Relational Programming Workshop}{September 9 2024}{Milan, Italy}
\acmMonth{9}
\acmArticle{x}
\copyrightyear{2024}
\acmYear{2024}
\acmDOI{}
\setcopyright{none}
\usepackage{fancyhdr}
\author{Evan Donahue}
\date{\today}
\title{Relational Reactive Programming: miniKanren for the Web}
\begin{document}

\email{evan.donahue@tc.u-tokyo.ac.jp}
\affiliation{%
\institution{University of Tokyo}
\country{Japan}
}

\settopmatter{printacmref=false}
\settopmatter{printfolios=true}
\renewcommand\footnotetextcopyrightpermission[1]{}
\pagestyle{fancy}
\fancyfoot{}
\fancyfoot[R]{miniKanren'24}
\fancypagestyle{firstfancy}{\fancyhead{}\fancyhead[R]{miniKanren'24}\fancyfoot{}}
\makeatletter
\let\@authorsaddresses\@empty
\makeatother

\begin{abstract}
Over the past decade, reactive frameworks and languages have become the dominant programming paradigm in front-end web development. In this paradigm, user actions change application state, and those changes propagate reactively to derived state and to the display, reducing the likelihood that various parts of the data model and user-facing view will become out of sync due to programmer error. In this paper, we explore the application of relational programming to the specification and synchronized evolution of model and view across time in response to user input. To that end, we present a reactive Javascript implementation of miniKanren and an integrated reactive programming model oriented towards the challenges of front-end web development. 
\end{abstract}

\maketitle
\thispagestyle{firstfancy}
\title[Goals as Constraints]{Goals as Constraints: Writing miniKanren Constraints in miniKanren}
\pagenumbering{gobble}
\section{Introduction}
\label{sec:org1ca575e}
The past ten years have seen a dramatic increase in the complexity of web applications, which in turn has led to a proliferation of languages and frameworks designed to simplify the development of web user interfaces. Over time, these languages and frameworks have tended to converge on a reactive, functional paradigm that emphasizes both the propagation of changes among mutually dependent elements of the data "model," as well as the automatic synchronization of the user-facing "view" with changes to that model. 

In this paper, we present a Javascript implementation of miniKanren with an integrated reactive engine for solving the problem of specifying the dynamic dependencies between model and view elements in the context of browser-based application development. Our objective is to understand the potential strengths and limitations of a relational approach to application development. To that end, we consider how relational techniques can simplify both the expression of state transition rules for arbitrary application state as well as the dynamic derivation of a synchronized view from an evolving model.

There are two key insights behind our approach. The first is that relational queries can often replace updates to derived datastructures as a means for expressing state transitions, leading to potentially useful programming patterns. Such modes of expression are enabled by the realization that the substitution data structure at the heart of miniKanren implicitly tracks a great deal of provenance information, which a reactive system can leverage to propagate changes to connected parts of the model. We therefore propose a system that treats the miniKanren substitution as a persistent, global data store for all application data, and computes transitions between application states by modifying the substitution. We achieve this by introducing the concept of reactive unification, which defines application state transitions in terms of transformations of the substitution. 

The second key insight is that the dynamic structural elements of the view can be modeled using miniKanren's nondeterministic search, with each returned answer corresponding to a unique view element. Modeled this way, failed unifications and other constraints can automatically prune elements of the view that depend on data that does not exist or that otherwise violate the constraints of the current application state. We use a reified version of the miniKanren search tree as a persistent source of identity for such view elements, which not only allows the expression of synchronized views, but also grants a degree of control over the view update policy.

The remainder of this paper is divided into four main sections. In Section \ref{sec:orgcc05784}, we describe a reactive system defined in miniKanren and give examples of its use. In Section \ref{sec:orge261bed}, we give a light formalization of reactive unification and related aspects of the reactive system in order to make intuitions regarding its functioning more precise. In Section \ref{sec:org835e65d}, we supply some implementation details regarding the most significant extensions to the underlying miniKanren engine. Section \ref{sec:org0451e76} discusses several lines of related research that intersect with the present approach. Finally, Section \ref{sec:org54b2acc} describes open challenges we have encountered to motivate future research as well as offers a larger vision of the potential for application development using miniKanren.

\section{Interface}
\label{sec:orgcc05784}
This section describes the usage of the system for specifying reactive interfaces. Although the idiosyncrasies of browsers and the document object model are of only tangential relevance to the presentation of the core reactive system, we nevertheless offer a simplified explanation of these concerns in order to lend concreteness to the subsequent discussion. We begin by discussing the Javascript embedding of miniKanren used in this system, the data model, and a simplified templating language used in our experiments to construct reactive interfaces. We then discuss in more depth the new reactive unification operator and its associated usage.

\subsection{miniKanren Embedding in Javascript}
\label{sec:org2fa5a45}
By way of illustrating the Javascript embedding of miniKanren used in this paper, we offer the following simple goal:

\lstset{frame=none,language=javascript,label= ,caption= ,captionpos=b,numbers=none}
\begin{lstlisting}
fresh((x,y) => [x.eq(y), conde(y.eq(1), y.eq(cons(1,2)))])
\end{lstlisting}

This goal introduces two fresh variables, \lstinline|x| and \lstinline|y|, unifies them, and then nondeterministically unifies \lstinline|y| with either \lstinline|1| or \lstinline|(1 . 2)|. Goals in this implementation are first-order, meaning that \lstinline|fresh|, like its subgoals, returns a concrete goal representation, rather than a stream of answers.

The \lstinline|fresh| function supplied by the miniKanren library accepts a function of arbitrary arity and calls it with the appropriate number of fresh variables. the Javascript syntax \lstinline|(x,y) => <expression>| evaluates to an anonymous function of two arguments, \lstinline|x| and \lstinline|y|, which returns the result of evaluating \lstinline|<expression>| with \lstinline|x|  and \lstinline|y|  appropriately bound. \lstinline|fresh| assumes that its function argument will return a goal, or as in this case an array of goals, which is interpreted as a conjunction of the contained subgoals \lstinline|x.eq(y)| and \lstinline|conde(y.eq(1), y.eq(cons(1,2)))|. 

\lstinline|conde| is a global variadic function that produces a disjunction of its argument goals, which may also be arrays of goals that signify conjunctions.

The use of \lstinline|cons| here deserves special mention, as Scheme-style cons lists are not natively a part of Javascript. The miniKanren library supplies functions such as \lstinline|cons| and \lstinline|list|, which produce Javascript objects containing "car" and "cdr" properties. These functions are supplied as a means to ease the expression of miniKanren programs, which traditionally make use of such operators. Although unification is defined more generally for arbitrary Javascript objects, as will be discussed in Section \ref{sec:orga5cdbd6}, it nevertheless reduces to the familiar procedure when confined to objects created with these functions.

\subsection{Model}
\label{sec:org31bbe76}
At the core of any reactive system is the fundamental data from which the interface is ultimately derived. This data may be fetched from a server or read from disk and it contains all information necessary for the operation of the application. In the context of user interfaces, this data is typically referred to as the "model." 

For the purposes of the current system, the model is a single, fully ground value as defined by the particular embedding of miniKanren. For the implementation used in this paper, the miniKanren embedding recognizes Javascript strings, numbers, and functions as atomic values, and Javascript objects and arrays, which are simple maps between string names (or numbers) and arbitrary values, as complex terms. 

While the model may be as simple as a single primitive, such as an integer counter, it may also be an arbitrarily complex, nested object. At each timestep, the current view is guaranteed to reflect values present in or derived from this model, and every user interaction that changes the application state can be defined in terms of an arbitrary update to this model using reactive unification.

\subsection{View Templates}
\label{sec:orgd9e4a10}
There are many solutions to the problem of specifying the relationship between a reactive data model and the user-facing interface, typically referred to as the "view." For the purposes of this presentation, we describe a simple s-expression-based templating language that uses Javascript arrays to specify HTML markup. While this templating language is somewhat orthogonal to the operation of the reactive miniKanren system, it will serve to make the presentation of the latter easier in the following sections.

HTML is a hierarchical markup language that specifies a tree of nodes, each containing zero or more children and zero or more attribute-value pairs. When the browser receives an HTML document, it parses the HTML string and reflects the resultant tree into the Javascript runtime as a tree of node objects via the Document Object Model (DOM) interface. The templating language described in this section foregoes the generation of HTML strings and directly constructs a document node object in Javascript according to the following informal syntax:

\begin{align*}
Template &\rightarrow String\\
Template &\rightarrow [Properties, Template, ...]\\
Properties &\rightarrow \{Name: Property, ...\}\\
Name &\rightarrow String\\
Property &\rightarrow String\\
\end{align*}

Strings are converted directly into text node objects and displayed as such. Arrays, denoted by square brackets, are converted into parent nodes with their tag name and other attributes specified by a Javascript literal Properties object mapping attribute names to values. Subsequent elements of the array are appended as children to the parent node.

Concretely, the following template constructs a paragraph tag (\lstinline|p|) with the id attribute "text" and a single child text node containing the string "lorem ipsum":\footnote{Property names such as "tagName" and "id" are defined by the DOM specification and passed mostly transparently through to the DOM API, with some exceptions for non-writeable properties such as "tagName," which must be handled specially.}

\lstset{frame=none,language=javascript,label= ,caption= ,captionpos=b,numbers=none}
\begin{lstlisting}
[{tagName: 'p', id: 'text'}, 'lorem ipsum']
\end{lstlisting}

\subsection{Reactive Templates}
\label{sec:org1d82dbf}
The static interface can be made reactive by replacing elements of the static syntax with miniKanren goal constructors. In particular, we add the syntactic rules:

\begin{align*}
Template &\rightarrow GoalConstructor\\
Property &\rightarrow LogicVar\\
Template &\rightarrow GoalConstructor\\
Property &\rightarrow LogicVar
\end{align*}

In place of any Template or Property, the user may place either a logic variable or a function of one argument. In the case of the logic variable, it must be bound in the substitution to a Template that will be rendered in its place. In the case of the function, it accepts one logic variable and returns a goal that binds that variable to a template. We will refer to the variables bound to view templates in these two cases as "view" variable. The bound views may incorporate application data by destructuring a global "model" variable that points to the root of the application model data, or any variable bound to some part of that model by other goals. 

Assuming the \lstinline|model| variable  was bound to the raw string "lorem ipsum," the following reactive template would reproduce the paragraph tag from the previous example by binding the view variable directly to the model variable using the \lstinline|eq| unification method on logic variables.

\lstset{frame=none,language=javascript,label= ,caption= ,captionpos=b,numbers=none}
\begin{lstlisting}
[{tagName: 'p', id: 'text'}, view => view.eq(model)]
\end{lstlisting}

As the model evolves over the course of the application, these view variables will be reactively rebound and the view automatically kept in sync with the model.

\subsection{Dynamic Reactive Templates}
\label{sec:org5814cc6}
When the mapping from model variables to view elements is one-to-one, the design of reactive systems is comparatively simple---all of the dependencies are present at the start and can therefore be specified at the source code level. The more challenging case is when dynamic model values that may change their structure over time, such as lists with varying numbers of elements, must be put into correspondence with a dynamic set of view elements. Reactive miniKanren addresses this problem through the use of nondeterminism. 

Goal constructors that return multiple answers generate multiple templates, which are rendered and inserted into the document as sibling elements. Hiding and showing a single view element and adding and removing a dynamic number of view elements are handled uniformly by controlling the number of answers returned by a goal constructor. The following template, for example, assumes that the model is a list of strings, such as \lstinline|('lorem' 'ipsum')| , and inserts an arbitrary number of paragraph elements containing those strings as body text:

\lstset{frame=none,language=javascript,label= ,caption= ,captionpos=b,numbers=none}
\begin{lstlisting}
view => fresh(text => [model.membero(text), view.eq(['p', text])])
\end{lstlisting}

The \lstinline|membero| instance method of the \lstinline|model| variable binds each element of the list bound to that logic variable to the logic variable passed in as an argument. The view variable is then bound nondeterministically to a paragraph element containing the list element as a child node.

It is important to note that all goal constructors implicitly run under the semantics of \lstinline|run*|, which is to say that all possible answers are exhausted. This means that the final set of rendered view elements is deterministic up to order. Moreover, because all answers are guaranteed to be returned, a depth-first search order is used in place of the traditional miniKanren interleaving search to add predictability to the order of outputs. More complex orderings are possible via an optional "order" variable argument to goal constructors, which can be bound to an item that explicitly defines the sort order for rendered views, as in the following example:

\lstset{frame=none,language=javascript,label= ,caption= ,captionpos=b,numbers=none}
\begin{lstlisting}
(view, order) => conde([view.eq('ipsum'), order.eq(2)],
                       [view.eq('lorem'), order.eq(1)])
\end{lstlisting}

This template generates two sibling text nodes in the correct "lorem ipsum" order despite the depth-first conde returning the "ipsum" node first. This is because the order variable has been bound in each answer to a value that override the default search order.\footnote{The default ordering function is an ascending order using the host language's own comparison function. The full implementation allows this ordering function itself to be overridden dynamically within miniKanren, although we omit further details of this API to simplify the exposition.}

This model of dynamic view rendering extends naturally to the recursive case where the number of nesting levels is not known in advance. The following recursive template unfolds an arbitrarily nested list and renders each string as a list element (\lstinline|li|) and each sublist as a nested unordered list (\lstinline|ul|) of list elements:

\lstset{frame=none,language=javascript,label= ,caption= ,captionpos=b,numbers=none}
\begin{lstlisting}
['ul',
 view => (function treeview(view, model) {
     return conde([model.isPairo().not(), view.eq(['li', model])],
                  [model.isPairo(),
                   view.eq(['li', ['ul',
                                   subview => fresh(submodel =>
                                       [model.membero(submodel),
                                        treeview(subview, submodel)])]])])}
         )(view, model)]
\end{lstlisting}

This template uses the recursive Javascript function \lstinline|treeview| to test the value bound to the \lstinline|model| variable. If it is a string, the view variable is bound to a list item with the string as its child. If the model is a list, the view is bound to a template containing a recursive goal constructor that nondeterministically binds sub elements of the list as models to subviews, which are added as children of the parent view.

\subsection{User Inputs \& State Change}
\label{sec:org1a8c77c}
Event handlers for user inputs are specified in much the same way as other reactive portions of the template. Every type of node in the DOM capable of generating events does so by calling all event listener functions registered to that node on a specific event type channel specified by a string. In this templating system, node properties that begin with the substring "on"---following the naming conventions of HTML for specifying event listeners---create reactive event listeners. Consider the following example:

\lstset{frame=none,language=javascript,label= ,caption= ,captionpos=b,numbers=none}
\begin{lstlisting}
[{tagName: 'div'},
 [{tagName: 'p'}, model],
 [{tagName: 'button', onclick: model.set('dolor sit amet')}]]
\end{lstlisting}

Assuming, as before, that the model consists of the simple string "lorem ipsum," the paragraph tag will display this text until the button is clicked, at which point the text in the model and correspondingly in the paragraph tag will be replaced by the next few words of the passage, "dolor sit amet." This is done with the \lstinline|set| logic variable method, which corresponds to the reactive unification operation described in Section \ref{sec:orgcc135a9}. For now, reactive unification can be thought of as roughly equivalent to mutable assignment. The string contents of the model variable are overwritten by the new string. Event handlers are specified as miniKanren goal constructors of local, lexically bound logic variables. 

\section{Formalization}
\label{sec:orge261bed}
This section offers a formalization of reactive unification and relevant underlying elements of the reactive system in order to make more precise some of the intuitions offered by the previous section. 

First, we define a relational reactive system as a 4-tuple of a substitution \(\mathcal{S}\), a model root variable \(\mathcal{M}\), a set of pairs of view variables \(\mathcal{V}\) and view goals \(\mathcal{G}\), and a set of update goals \(\mathcal{U}\):

\begin{align*}
\langle\mathcal{S}, \mathcal{M}, \{\langle\mathcal{V}, \mathcal{G}\rangle\}, \{\mathcal{U}\}\rangle
\end{align*}

The substitution \(\mathcal{S}\) contains all of the application data. Model variable \(\mathcal{M}\) is bound in \(\mathcal{S}\) to the root value of the application data model, and reifying \(\mathcal{M}\) within \(\mathcal{S}\) yields precisely the current state of the application. 

Each view variable \(v \in \mathcal{V}\) corresponds, in the DOM context, to a set of nodes or to a node attribute, although the details of how these nodes are displayed is orthogonal to the operation of the reactive system. Abstractly, the set of reified assignments to each view variable defines the state of the application's view at any given time, regardless of how those assignments are displayed by a specific interface technology. Each such assignment can be derived by running the view variable's corresponding view goal \(g \in \mathcal{G}\) on \(\mathcal{S}\) and reifying \(v\) in the resulting substitutions contained in each answer. 

State transitions are defined as a union of answers produced by running a single update goal \(u \in \mathcal{U}\) on \(\mathcal{S}\), collecting the resulting set of reactive unifications, and applying them to \(\mathcal{S}\). 

The remainder of this section explicates the nature of these components by walking through the lifecycle of a reactive system while expanding as needed on important details relevant to each step.

\subsection{Substitution Normal Form}
\label{sec:orgdd6d037}
The first stage in the lifecycle of a reactive application involves loading the model data from the disk or network into the reactive system. One of the two central ideas in this presentation is that all model data in the reactive system is stored in a miniKanren substitution data structure. Subsequently, goals will be run to bind view variables to data contained in this substitution and events will fire that modify the substitution from one timestep to the next through reactive unification. However, at no point during the lifecycle of the application until data is serialized back to the disk or network is any application data held in any form outside of the substitution. As a result, we describe several modifications to the structure and usage of the substitution datastructure to enable its persistence across timesteps and use in specifying fine-grained reactivity, starting with the definition of a normal form. 

Reactive systems require the ability to recompute data in any part of the reactive model when its dependencies are modified. Updating a single value may cause goals to be re-run, other values to be recomputed, or elements of the view to be redrawn. Consequently, some notion of identity is required to allow views to specify the model values on which they depend. We use logic variables as a form of identity by requiring every atomic piece of data in the substitution to be associated with a single, unique logic variable. Given the substitution for an incrementing counter, \(x \mapsto 0\), any elements of the view that must display the counter can rely on being able to reify the variable \(x\), as defined by object identity within the host language, to determine the current count to display. Likewise, incrementing the counter will yield  \(x \mapsto 1\), while preserving \(x\) as the unique identity of the counter data over the lifetime of the reactive application. This allows any part of the model to be changed in arbitrary ways through reactive unification by specifying, on the left-hand side, the variable corresponding to that single, unique part of the data model.

In order to guarantee that each piece of potentially mutable data corresponds to a unique logic variable identifier, we start by normalizing the application data to achieve this condition. We define the normal form of the substitution as follows: 

\begin{enumerate}
\item Every logic variable that appears anywhere in \(\mathcal{S}\) must be bound in \(\mathcal{S}\) to a non-logic variable
\item Every complex term, such as a pair or any other container objects defined by the implementation, must only contain logic variables
\item Starting from \(\mathcal{M}\), the logic variables in \(\mathcal{S}\) must form a tree in the sense that if two variables \(u\) and \(v\) share a common descendant \(w\), then \(v\) must also be a descendant of \(u\) or vice versa. No variables are shared between distinct complex terms, and there are no cycles.
\end{enumerate}

With these properties, any single piece of data can be changed arbitrarily without affecting any other piece. Complex terms, moreover, can be modified without losing reference to the data they contained.

Normalizing the initial application data corresponds roughly to walking the model value and inserting each datum into the substitution bound to a new logic variable. This procedure is a special case of the more general reactive unification procedure described in Section \ref{sec:org30f8409}, corresponding to reactively unifying the entire model with a fresh variable in an empty substitution. 

\subsection{Reactive Unification}
\label{sec:orgcc135a9}
One of the central contributions of this paper is the reactive unification operator, which introduces into miniKanren a limited notion of time sufficient to specify a transition relation between an application state and its successor state. In its simplest usage, exemplified in \ref{sec:org1a8c77c}, reactive unification can be thought of as assignment at the next timestep of the value of the right-hand term to the variable on the left-hand side. To that end, reactive unification is asymmetric, and the left-hand side must be a logic variable that is either bound in \(\mathcal{S}\) or else will eventually be unified with such a logic variable. The right hand side may be any valid miniKanren value.

The main advantage of reactive unification is that it allows miniKanren queries to simultaneously generate views as well as specify updates. Because atomic values such as numbers and strings are neither created nor destroyed by pure relations, any such values that appear in the finally computed view as a result of computations applied to the model must point, by indirection through the substitution's bindings, to the same atomic objects present in the data model. This makes it trivial to update atomic values directly no matter where in the computation they originated, and no matter how many layers of computation intervene between input and output, by simply applying reactive unification to any variable bound to those values at any point in the computation. 

The precise semantics of reactive unification are an active area of research at the time of writing. However, in this section we outline a number of candidate properties that enable potentially useful programming patterns and properties in practice. Our goal in enumerating these properties is to arrive at a semantic model that can succinctly and reliably express the state transitions from one timestep to the next while minimizing the cognitive burden on the programmer. Much of the promise of the relational approach to reactive interface design hinges the ability of this operator to express bidirectional update logic aided by the relationality of miniKanren's design. This list is inspired by similar enumerations of lens properties in the bidirectional programming literature, albeit much less developed \cite{fischer2015clear}.

\subsubsection{Relational PutGet Law}
\label{sec:org40c295d}
The PutGet law for functional lenses states that the value written to a source should be the value returned by subsequent reads \cite{fischer2015clear}. We adapt this law for reactive unification in order to preserve the reorderability of relations---an important property in miniKanren---as follows:

\begin{align*}
x_{t+1} \equiv y_t \implies \hat{x}_{t+1} = \hat{y}_t
\end{align*}

Here, \(x_{t+1} \equiv y_t\) signifies reactive unification of variable \(x\) at timestep \(t+1\) with the value of variable \(y\) at timestep \(t\). \(\hat{x}_{t+1}\) and  \(\hat{y_t}\) are the reified values of logic variables \(x\) and \(y\) at times \(t+1\) and \(t\), respectively.\footnote{We disallow free logic variables in the model and view, and so assume for present purposes that all reified values are fully bound. Consequently, simple equality in the host language is sufficient to define equivalence of reified terms without needing to resort to alpha equivalence.} This states that the reified value of a variable should be equal to the reified value of the variable or value to which it was set in the previous timestep using reactive unification.

This law, interpreted in a relational context, has several additional implications, such as a significant weakening of the PutPut law. PutPut states that for two consecutive writes to a source, the second write should overwrite the first and be returned by subsequent reads. However, in the relational context the question of which write is second is indeterminate, as goals may be reordered. This yields the weakened form of PutPut:

\begin{align*}
x_{t+1} \equiv y_t \land x_{t+1} \equiv z_t \implies \hat{y}_t = \hat{z}_t
\end{align*}

Which states that if a variable is reactively unified twice, the reified values of the terms to which it is set must be equal. Otherwise, reactive unification fails and the term is unchanged.

An additional implication is that updates must be temporally stratified. Consider the simultaneous swapping of two variables:

\begin{align*}
x_{t+1} \equiv y_t \land y_{t+1} \equiv x_t \implies \hat{x}_{t+1} = \hat{y}_t \land \hat{y}_{t+1} = \hat{x}_t
\end{align*}

If updates happened within a timestep, this would impose an ordering on the updates at odds with the constraints of relational goal reorderability. As such, values at one timestep must be fixed from the perspective of updates in the subsequent timestep.

\subsubsection{Equivalence Class Assignment}
\label{sec:orge9f1d21}
This property states that all variables that belong to the same equivalence class as defined by mutual unification should be set to the same value at the next timestep if any of them is set through reactive unification. More precisely:

\begin{align*}
x_t \equiv y_t \land x_{t+1} \equiv z_t \implies \hat{x}_{t+1} = \hat{y}_{t+1} = \hat{z}_{t}
\end{align*}

Intuitively, some version of this property is a practical necessity. Goal constructors only receive direct references to the model variable bound to the root value of the model. Any reactive unifications that target subterms of the model must do so indirectly by reactively unifying destructuring variables bound to subterms of the model. Without any version of this property, the only useful left-hand argument for reactive unification would be the root model variable itself, requiring programmers to reconstruct the entire model manually at each timestep in order to update it. With this property, we aim for a more mutational metaphor according to which subterms can be set directly. For example, consider the case of replacing the head of a list \lstinline|x| with the string "lorem ipsum":

\lstset{frame=none,language=javascript,label= ,caption= ,captionpos=b,numbers=none}
\begin{lstlisting}
fresh((y, z) => [x.eq(cons(y, z)), y.set('lorem ipsum')])
\end{lstlisting}

Because we do not have direct access to the variable bound to the current head of the list, we must bind a fresh variable \lstinline|y| and reactively unify that variable with the new string. In order for this operation to make sense, we must be sure that the variable to which \lstinline|y| is ultimately bound is updated as well, as it is this original variable that stores the ground truth information about its value.

\subsubsection{Deterministic Assignment}
\label{sec:orgb78a6d3}
As an arbitrary miniKanren goal, update goals may return multiple answers. We define the total reactive unification relating one application state to its successor state as the conjunction of all reactive unifications in all answers \(A\) returned by a given update goal.

\begin{align*}
\forall y,a\in A\ \ \ \ x^a_{t+1} \equiv y^a_t \implies \hat{x}_{t+1} = \hat{y}^a_t
\end{align*}

In other words, at the next timestep, the reified value of variable \(x\) should be equal to the reified values of all variables \(y\) bound to \(x\) in any answer. This in turn implies that \(y\) must reify to the same value in all states in which it is bound to \(x\) lest it violate PutGet.

This definition is one way to reduce the ambiguity of the otherwise unspecified behavior as to how to treat multiple, non-deterministic answers returned by an update goal. Moreover, it has the desirable property that it allows programmers to leverage miniKanren's search capabilities and reuse existing relations used to derive data in the forward direction to propagate changes backwards without writing a separate class of update relations. For example, consider the case of mapping over a list and duplicating each element:

\lstset{frame=none,language=javascript,label= ,caption= ,captionpos=b,numbers=none}
\begin{lstlisting}
fresh(y => [x.membero(y), y.set(cons(y, y))])
\end{lstlisting}

By leveraging search to nondeterministically reactively unify \(y\), it is possible to avoid the need to write a dedicated deterministic relation that walks the list and reactively assigns new values. Moreover, we observe that because the substitution preserves the provenance of all atomic data within the system, it is possible to set any atomic value from any reference to that value no matter how many intervening computations have processed that data. This gives the programmer significant leverage in specifying state transitions without writing dedicated update relations. This enables techniques such as relational joins, where answers may contain data combinatorially composed from a variety of sources (such as by conjoining several \lstinline|membero| relations) and reactive unification will be able to overwrite any of those values by tracing them to their respective sources in the substitution.

\subsubsection{Recursive Stratification}
\label{sec:org37f2c65}
One final property of potential value is recursive stratification. This property technically conflicts with earlier properties, such as the relational PutGet law, in that it requires cases that would generate failures in the PutGet law to instead return valid answers. Rather than complicate the presentation of earlier properties, we describe it here as an exception.

The intuition behind this property is that if we assign a value to the tail of a list, and then assign a value to the tail of the tail, we would like both operations to succeed rather than fail due to conflicts. Specifically, we would like to support the pattern of using deterministic assignment to modify the structure of a complex structure such as a list. Consider the following code which removes all instances of "lorem" from a list:

\lstset{frame=none,language=javascript,label= ,caption= ,captionpos=b,numbers=none}
\begin{lstlisting}
fresh((cdr, tail) => [tail.eq(cons('lorem', cdr)), model.tailo(tail), tail.set(cdr)])
\end{lstlisting}

\lstinline|tail| is bound to each cons pair headed by "lorem," and so by setting those cons pairs to their cdrs, the intent is to locally modify the list to remove each matching pair. Without the present property, if there is more than a single "lorem" in the list, this goal will amount to setting sublists of the tail to conflicting values, resulting in failure. With special handling, however, it is possible to make the above goal work in accordance with the intuition given. This allows "pointers" to structural elements of the model to be bound to logic variables and locally set through reactive unification to change the structure of the model in a somewhat controlled way. 

This mutable metaphor avoids some of the traditional problems of such mutability. For instance, structure cannot be shared between model variables due to the normalization described in Section \ref{sec:orgdd6d037}, so local changes are somewhat protected from interference by other reactive goals. Moreover, because changes are still temporally stratified, there is no risk of changing the list one is iterating over as one iterates. However, further experience is necessary to determine the limits or liabilities of this property.

\section{Implementation}
\label{sec:org835e65d}
This section offers some technical details pertaining to reactive unification and the view updating strategy used by the reactive system. 

\subsection{Reactive Unification}
\label{sec:org30f8409}
\subsubsection{Revisiting Normal Unification}
\label{sec:orga5cdbd6}
Before we can define reactive unification, it is necessary to modify ordinary unification slightly so that operations performed by reactive unification remain consistent with the properties proposed in Section \ref{sec:orgcc135a9}. When miniKanren is used for search, the substitution is typically immutable. In this case, there is no practical difference between two variables \emph{happening} to be equal and being \emph{constrained} to be so. Consider, for example, the substitution \(x \mapsto 1 \land y \mapsto 1\). Unifying \(x\) and \(y\) in this substitution, in most implementations, yields the same substitution. The information that \(x\) and \(y\) have been unified is thrown away because for all intents and purposes, it is enough to know that \(x\) and \(y\) have the same value, regardless of that value's provenance. 

To account for reactive unification, we must record the unification between \(x\) and \(y\) even if the two are already equal, so that if we later reactively update \(x\), we can ensure that \(y\) is appropriately updated as well. We capture this information by ensuring that both \(x\) and \(y\) walk to the same variable. For instance, we could remove the binding for \(y\) and rebind it to \(x\), yielding the substitution \(x \mapsto 1 \land y \mapsto x\). To make this intuition more precise, we can define the equivalence class of all mutually unified variables in a given substitution as the set of variables that share a common descendant, with ancestry defined by the walk procedure:

\begin{align*}
x \equiv y \implies \exists z \ \  descendant(x,z) \land descendant(y,z)
\end{align*}

Where the \(descendant\) relation is recursively defined as:

\begin{align*}
descendant(x,y) \implies x = y \lor \exists z\ \ (x \rightarrow z) \in\mathcal{S} \land descendant(z,y)
\end{align*}

We can then modify non-reactive unification to guarantee that each set of mutually unified variables contains one member that is the common descendant of all members of the set. This can be achieved by, each time two variables are unified, replacing the binding of one unified variable with a binding to the other variable, after confirming that they do not conflict. See Appendix \ref{sec:org15bb66e} for a full implementation. 

\subsubsection{Implementing Reactive Updates}
\label{sec:orgb8328a2}
With normal unification so defined, we can proceed to define reactive unification. Reactive unification begins with \(\mathcal{S}\) and a stream of answers \(\mathcal{A}\) produced by the update goal. Each answer \(a \in \mathcal{A}\) contains some number of stored reactive unification goals \(r \in a\). Each reactive unification goal \(r\) possesses a left-hand side variable to which the right-hand side value will be written. 

The update goal is not applied directly to \(\mathcal{S}\), but to an \(\mathcal{S}'\) that is the result of applying all of the goals contained in the view tree along the path from the root to the leaf containing the update goal. As such, reactive unification first generates a patch by calculating the diff entailed by the set of reactive unifications \(r\) applied to \(\mathcal{S}'\). The patch takes the form of a list of cons pairs in which the left-hand side is a variable bound in \(\mathcal{S}\) and the right-hand side is a ground value. This patch is then applied to \(\mathcal{S}\) to calculate the successor state. 

In a simple implementation, the left and right hand sides of each reactive unification can simply be walked and reified, respectively, in their corresponding answer state. In order to support the recursive stratification property from Section \ref{sec:org37f2c65}, however, an additional step is required. Specifically, the reification procedure must be extended such that when reification encounters a variable that is set directly by another reactive update, it must abandon further reification within the answer substitution and proceed with the reified results of that update. Updates that are descendants of other updates are subsequently discarded, as they have already been incorporated into ancestor updates.

Applying the patch is a simple unification-like procedure in which variables and patch values are walked in tandem, with the latter overwriting the former as applicable.

\subsection{View Tree Updating}
\label{sec:org7bc41cd}
This section details the implementation of a strategy for keeping stateful view elements in sync with changes to the model at each timestep. A naive implementation might rebuild the entire view at each timestep and replace the previous one completely. Performance concerns aside, this approach encounters trouble in the browser context due to the loss of implicit state, such as the currently focused text field or scrollbar offset, which can harm the user experience. As such, a strategy is needed to update the view piecewise in order to preserve implicit state in subviews unaffected by the most recent model changes.

One common approach to solving this problem is known as "virtual DOM," popularized by libraries such as React \cite{facebook2024react}. In virtual DOM-based approaches, a representation of the view is often rebuilt or modified at each timestep and compared with the representation at the previous timestep through a diffing procedure that results in a patch that can be applied to the live document. One common challenge of such approaches is that computing a minimal diff between two view trees is a computationally complex problem \cite{bille2005survey}, which can result in poor run times even in simple cases. 

Practical virtual DOM libraries have tended to address this time complexity challenge through the use of a combination of heuristics that cover common practical operations and "keyed" iteration, which provides syntax for programmers to compute unique indices for dynamic elements that the diff algorithm can use to establish identity for elements of the view for which the data can support unique derived indices.

\subsubsection{Persistent Search Tree}
\label{sec:org32d683a}
Reactive miniKanren lends itself to an interesting approach to the problem of synchronizing the view tree with the model, which, while its performance characteristics have yet to be tested in a practical environment, nevertheless possesses interesting properties that bear further investigation. 

Reactive miniKanren maintains a persistent tree of dynamic view elements and attributes, which corresponds roughly to the structure of the template that generated it. We will refer to as the "view tree." When \(\mathcal{S}\) is modified through reactive unification, its successor, \(\mathcal{S}'\), is passed down the branches of this tree, which re-run goals and update view elements as needed. In particular, goal constructor templates give rise to particularly interesting subtrees in that these subtrees are isomorphic to the structure of the miniKanren search performed by these goal constructors. Consider the following example of a pair of sibling text nodes containing the text "ipsum" and "dolor," respectively, as produced by the following goal constructor applied to \(\mathcal{M}\), \lstinline|("ipsum" "dolor")|:

\lstset{frame=none,language=javascript,label= ,caption= ,captionpos=b,numbers=none}
\begin{lstlisting}
view => model.membero(view);
\end{lstlisting}

\begin{figure}[htbp]
\centering
\includegraphics[width=.9\linewidth]{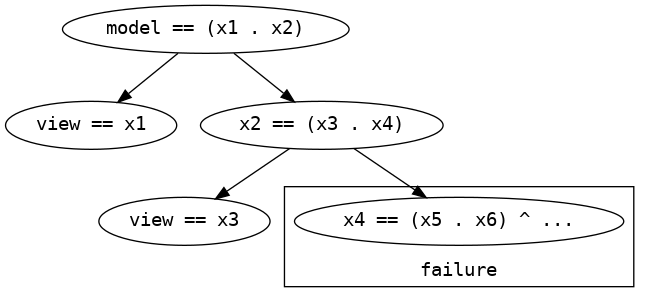}
\caption{\label{fig:org4468b20}
View tree generated by membero applied to a cons list}
\end{figure}

Because the list has two elements, \lstinline|membero| will succeed twice at destructuring the list and binding the view to its first and second elements before failing to destructure the empty tail of the list. Reactive miniKanren will therefore render this search as the view subtree in Figure \ref{fig:org4468b20}.

To build this tree representation, reactive miniKanren threads an additional value through the search procedure alongside the substitution that corresponds to a conjunction of all unifications and constraints seen to this point in the search. When the search encounters a \lstinline|conde|, it creates a branching node in the view tree, as in Figure \ref{fig:org4468b20}, containing the first-order conjunction of unifications and constraints seen to this point in the search. It then discards these conjuncts and proceeds with the trivial \lstinline|succeed| goal as its new conjunction. Likewise, when successful answers or failures are encountered, nodes are created containing the new current conjunction. 

This tree is isomorphic to the miniKanren search, and can be viewed as a first-order representation of that search. The top layer of Figure \ref{fig:org4468b20} , for example, can be read as destructuring the head of \(\mathcal{M}\) and then, in the second layer, nondeterministically both binding \(x1\) to the first element, "ipsum" and destructuring the tail of \(\mathcal{M}\). The third layer of the tree then represents the nondeterministic binding of \(x2\) to the second element of \(\mathcal{M}\) and the failed attempt to destructure the subsequent tail. This isomorphism allows reactive miniKanren to re-run its search with a new \(\mathcal{S}'\) by passing \(\mathcal{S}'\) down the branches of this tree and re-applying the stored goals at each node. 

The advantage of maintaining and modifying a representation of the search over simply re-running the search in toto is that the first-order search tree representation can store additional information about how its answers are bound to the live view elements that would be lost if the search information was discarded. Rather than producing a new representation of the view and performing potentially expensive diffs, nodes in this tree that transition from failing to succeeding states or vice versa can directly add elements to or remove them from the live DOM. Each answer's unique position as a node in this view tree, therefore, can be viewed as an implicit "key" in the above sense of keyed iteration that establishes a unique identity that is not dependent on deriving a unique identifier from the data. Because even failures possess such unique positions in the search tree, they preserve their implicit order in the stream of answers and can insert their corresponding view elements at the correct sibling position when they become non-failing, which is information that would be lost in a postprocessing of the answer stream via a diff-based method.

This approach, however, has both strengths and weaknesses due to the fact that the "keys" supplied by the position in the search tree are assigned without regard to the intended usage pattern of the underlying model. For example, the strategy described above excels in cases such as filtering a list according to a search term, which changes which nodes are displayed but does not change the relationship between a given node and its contents. The first node will remain bound to "ipsum" even as it is hidden and shown in response to changes in the filter term. Nodes not affected by visibility changes can remain as they are in the DOM. 

This model is much weaker, however, when it comes to cases such as insertion into a list. If an element is inserted at the beginning position of a list that was naively traversed with \lstinline|membero|, then the first, second, and all of the rest of the answers returned by the goal will have changed their value, and all nodes will have to be modified even though a more efficient strategy, from the perspective of the DOM, may have been to simply insert one node at the front of the list. Using Figure \ref{fig:org4468b20} as an example, if "lorem" was inserted at the head of the list, then due to the mutable metaphor through which reactive unification operates, the entire list would be overwritten in place by new values. Hence, \(x1\) becomes "lorem" while \(x2\) becomes "ipsum" and so on. Because key identity is derived from position in the search tree, every node will have to be modified in order to bring the view into synchronicity with the model.

To address such shortcomings, we observe that because we are using position in the search tree as a form of identity, we can control the type of key strategy used by the reactive system by changing the dynamics of the search tree. In the case of lists of sibling view nodes, this means that changing the list representation and associated traversal strategy can result in different keying strategies with different performance characteristics.

\subsubsection{Insertion Lists}
\label{sec:orgb9464d1}
We introduce here the concept of "insertion lists," which are conceptually related to techniques such as difference lists in which the underlying representation and operations of list structures are modified, with resultant changes to performance and behavior \cite{hughes1986novel}. In particular, an insertion list is a list backed by a binary tree as the underlying representation. List order is defined by an in-order traversal of the tree. \lstinline|membero| can be implemented for such lists, then, as a simple tree traversal:

\lstset{frame=none,language=javascript,label= ,caption= ,captionpos=b,numbers=none}
\begin{lstlisting}
function imembero(xs, x) {
    return conde([xs.pairo().noto(), xs.eq(x)],
                 [fresh((a,b) => [xs.eq(cons(a,b)),
                                  conde(imembero(a, x), imembero(b, x))])])}
\end{lstlisting}

\begin{figure}[htbp]
\centering
\includegraphics[width=.9\linewidth]{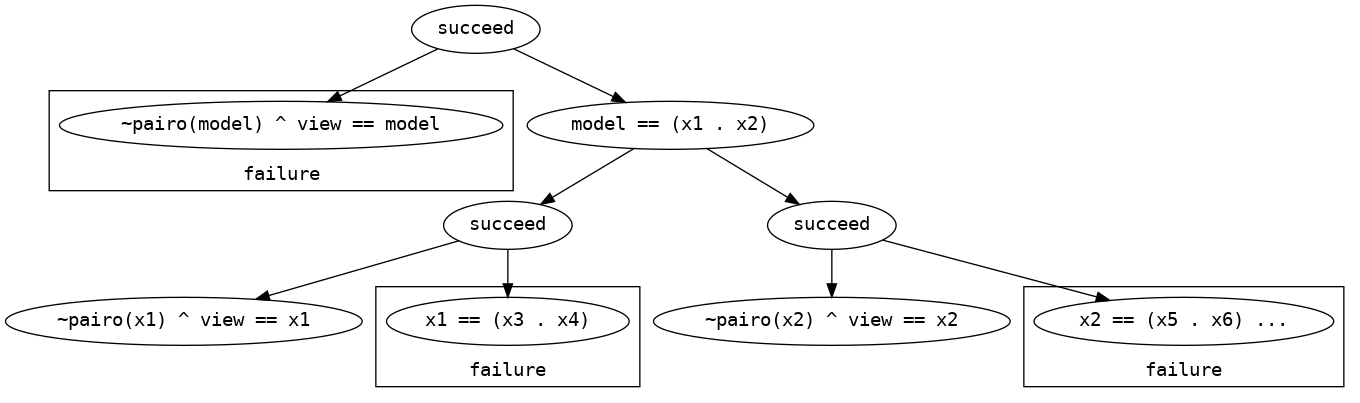}
\caption{\label{fig:org672f01a}
View tree generated by imembero applied to an insertion list}
\end{figure}

When used as a drop-in replacement for ordinary cons lists and \lstinline|membero|, the resulting view remains the same. However, consider the view tree associated with the goal \lstinline|imembero(model, view)| when the model is the insertion list \lstinline|("ipsum" . "dolor")|, as depicted in Figure \ref{fig:org672f01a}.

Reading this diagram from top to bottom, the second level corresponds to the nondeterministic choice of whether \(\mathcal{M}\) is a displayable non-pair, signified by the failing left hand check, or the right hand check, which destructures the pair. The fourth level likewise nondeterministically checks that the two children of \(\mathcal{M}\), "ipsum" corresponding to \(x1\) and "dolor" corresponding to \(x2\), are displayable non-pairs and binds them to the view variable. 

If we now define the insertion procedure as replacing a leaf of the binary tree representation with a binary tree of depth 1, inserting "lorem" at the beginning of the list yields \lstinline|(("lorem" . "ipsum") . "dolor")|. After reactive unification, this means that \(x1\), previously bound to "ipsum" is now bound to the binary tree \lstinline|("lorem" . "ipsum")|. The view tree will therefore swap the failure statuses of the left-hand grandchildren and further expand the previously failing pair node to account for the new substructure, binding \(x3\) and \(x4\) in the process. 

Note, however, that \(x2\) remains unchanged by reactive unification and therefore its corresponding view node does not need to modify the DOM. Likewise, all subsequent elements would be unaffected by this insertion, yielding the same final result as in the \lstinline|membero| case, but doing so by removing only one node and replacing it with two new ones, as opposed to changing every node in the list. Note too that the insertion list retains the same rough performance characteristics as the cons list in cases such as list filtering, albeit with the additional base overhead required to perform efficient insertions. It is possible that other list representations or traversal strategies could result in other useful performance profiles. While it is ultimately an empirical question whether these strategies result in net performance gains in a real-world application, because DOM operations tend to be expensive relative to the execution of pure Javascript, we argue that such properties represent a promising line of future inquiry.

One important caveat is that as a result of the isomorphism between the search tree and the view tree, any \lstinline|conde| expressions that serve only to select between two possible values for the same view element will produce redundant view elements for each branch of the \lstinline|conde| clause, potentially impacting performance. This is analogous to the problem in normal miniKanren of using \lstinline|conde| to write constraints, such as checking that a list is a proper list, and paying the performance penalty when that relation generates useless list structures when its arguments are free. It is likely, however, that this problem can be solved in a manner analogous to that of normal miniKanren through the use of non-generative disjunctive constraints, so this may not ultimately pose a great problem \cite{donahue2023constraints}.

\section{Related Work}
\label{sec:org0451e76}
The work presented in this paper lies at the intersection of three distinct lines of research, the first of which is research on functional reactive programming. Functional reactive programming originated as a paradigm for building animations by composing time varying values using the tools of functional programming \cite{elliott1997functional}. However, it has since evolved to incorporate a wide range of loosely connected techniques in a range of fields including distributed systems \cite{webster2022using}, real-time systems \cite{tasnim2023future}, and especially user interface programming \cite{czaplicki2013asynchronous,shukla2022bridging}.

Due to the forward-directional nature of functional evaluation, this research has historically placed less of an emphasis on "multidirectionality," or propagating user inputs "backwards" to update the model \cite{bainomugisha2013survey}. This omission has remained largely true of many of the contemporary systems in wide use. The present work, by contrast, has emphasized multidirectionality resulting from miniKanren's relational nature.

The second related area of research is that of bidirectional programming, such as research on optics as well as the closely related view update problem from the database literature \cite{bohannon2006relational,zhang2023contract}. Approaches such as lenses compositionally define get and set functions in parallel so that a program can transform a small piece of a larger, immutable data structure without needing to concern itself with reintegrating its changes into some deeply nested part of the original structure. The relationship between the current work and other bidirectional programming techniques warrants further investigation, and is discussed in more detail in Section \ref{sec:org54b2acc}. 

The final area of related research is that of adapting logic programming techniques to reactive application development. \citet{van2022florence}, for instance, adapts logic programming ideas for querying event streams, which is a significant component of FRP that has been left out of this presentation, but may represent a productive future direction for this research. There has also been research on using logic programming techniques to define application state transitions, focusing on state transitions that involve solving complex constraint problems to which logic programming techniques are well adapted, which may be an interesting paradigm to explore further \cite{kowalski2015reactive,kowalski2023combining}. 

\section{Future Work}
\label{sec:org54b2acc}
As the work presented in this paper is very preliminary, there are many natural avenues for future research. One such avenue is the exploration of fine-grained dependency analysis. One benefit of reactive systems that track fine-grained dependencies is that they can often avoid unnecessary computation when only a small part of the model changes. We observe that, due to the way in which we have defined the expansion of view trees, fresh variables can only be introduced the first time a goal is expanded. Afterwards, only unifications and other constraints are left in the nodes of the view tree. This means that, for any given subtree of the view tree that has been completely expanded, it is possible to determine exactly on which logic variables it depends. This information could, in principle, be exploited to avoid re-running unifications and constraints, which are more expensive even than they usually are in miniKanren due to the extra information unifications must store for the benefit of reactive unification. 

Regarding bidirectionality in particular, there are many interesting overlaps between the present work and past work on lenses and database views. This topic deserves a more expansive analysis, but provisionally we observe that the mutable metaphor by which we defined reactive unification seems to have the effect of encouraging the programmer to include "pointer" variables in views generated by queries, such as by binding the cons pairs of a list and later reactively unifying those pairs with their own cdrs to remove them from a source list. While this allows the programmer to be more explicit about changes in some cases where a view introduces ambiguity, such as inserting into a filtered view into a list, and is also without some of the drawbacks of traditional mutable patterns due to the strict temporal stratification of updates, it remains to be seen how these facilities will fare in the face of more complex practical problems. 

One additional avenue of potential research lies in reclaiming some of the nondeterminism that has been abandoned in this presentation for the purpose of precise programmer control over the system. Regarding nondeterminism in the model, consider the case of a negation relation: 

\begin{align*}
(x \equiv true \land y \equiv false) \lor (x \equiv false \land y \equiv true)
\end{align*}

Our presentation thus far has assumed a forward directional derivation from model to view. Assume, for instance, that \(x\) is a persistent part of the model data and \(y\) is therefore derived. In such a case, we could reactively unify \(x\) to rederive \(y\), but reactively unifying \(y\) would have no effect, even though it would be possible to determine, in theory, that such an update required the subsequent update of \(x\). The case becomes even more interesting when multiple possible update solutions exist, such as updating one summand in a sum, as this ambiguity may feed back into the design of the interface itself by using search to drive user queries for additional constraints. 

As discussed in Section \ref{sec:org1d82dbf}, goal constructors currently assume the semantics of \lstinline|run*|, leading to nondeterminism in the interface only with respect to order. However, a goal constructor with the semantics of \lstinline|run-1|, which returned only one satisfying answer, might prove interesting in that in this case the contents of the interface itself would be nondeterministic, allowing programmers to specify a space of possible interfaces and allowing the constraints to select the final user-facing result. 

Longer term, this work aims to be a small step towards a much larger goal of turning miniKanren's synthesis abilities towards the task of whole application synthesis. It would therefore be interesting to consider combining work on goal synthesis \cite{joshi2021metakanren} with past work on generating interface specifications \cite{kosarev2022declarative} to advance the problem of interactive application synthesis.

\section{Conclusion}
\label{sec:orgdde93ce}
In this paper, we have described a reactive implementation of miniKanren for building web application interfaces. Our focus in this implementation has been on leveraging miniKanren's pure relationality as a tool for solving a version of the view update problem as it commonly arises in complex event-driven interface design, and on repurposing miniKanren's nondeterminism to create efficient, dynamic views. We found that miniKanren lent itself well to these problems, and offers a promising avenue for future research in this area. In future work, we intend to continue allowing experience with building reactive interfaces to guide the development of a more complete model of relational reactive programming.

\section{Acknowledgments}
\label{sec:orgba554e5}
We thank Will Byrd for discussions of early versions of this idea. We also thank the anonymous reviewers for their suggestions, as well as Tokyo College at the University of Tokyo and the Digital Democracies Institute at Simon Fraser University for their support.

\bibliography{/home/apocalypsemystic/core/knowledgebase/cite.bib}
\bibliographystyle{ACM-Reference-Format}
\pagebreak
\appendix
\section{Normal Unification}
\label{sec:org15bb66e}
The below implementation of unification differs from the usual implementation in that, if two variables are unified, one must be bound to some descendant variable of the other (or to that variable itself) after confirming that the two are unifiable.

Unification here is an instance method of the substitution, which is bound to \lstinline|this|. 
\lstinline|walk_binding|
is analogous to normal \lstinline|walk|, but returns the final pair of bound variable and value rather than just the value (or a pair of variables or values if the walked term is a free variable or ground term). \lstinline|extend| extends the substitution without further processing. 

\lstinline|primitive| returns true if a ground term is considered a primitive within the host language, and non primitive terms (such as but not limited to Scheme-style cons lists and plain objects) are defined so as to unify if their overlapping properties unify. Properties not held in common between two objects are not considered for unification. This definition has advantages of programmer economy of expression when destructuring complex objects with narrow unifications, but we are still evaluating the potential ramifications of this definition. 

\lstset{frame=none,language=javascript,label= ,caption={Provenance-preserving Unification},captionpos=b,numbers=none}
\begin{lstlisting}
unify(x_var, y_var) 
    let x, y
    ({car: x_var, cdr: x} = this.walk_binding(x_var))
    ({car: y_var, cdr: y} = this.walk_binding(y_var))
    if (x === y) {
        if (x instanceof LVar || x_var === y_var) return this
        else if (x_var instanceof LVar) return this.extend(x_var, y_var)
        else return this.extend(y_var, x_var) }
    if (x instanceof LVar) return this.extend(x, y_var)
    if (y instanceof LVar) return this.extend(y, x_var)
    if (primitive(x) || primitive(y)) return failure
    let s = this
    for (let k of Object.keys(x).filter(k => Object.hasOwn(y, k))) {
        s = s.unify(x[k], y[k])
        if (s === failure) return failure }
    return s
\end{lstlisting}
\section{TodoMVC}
\label{sec:orge4eee52}
This section contains a complete implementation of a slightly simplified\footnote{The current implementation does not support numeric operations, and so the numeric counter view was omitted.} version of TodoMVC\cite{todomvc2024}, a standard UI "benchmark" that has been implemented in many well known Javascript frameworks in order to facilitate direct comparisons. The implementation takes the form of a function, \lstinline|template|, that accepts a model variable and returns the template. Due to the lexically scoped nature of miniKanren and reactive miniKanren templates, normal functional abstraction suffices for composing view templates.

\lstset{frame=none,language=javascript,label= ,caption={Complete implementation of TodoMVC},captionpos=b,numbers=none}
\begin{lstlisting}
function template(m) {
    return [{tagName: 'section', className: 'todoapp'},
            [{tagName: 'header', className: 'header'},
             ['h1', 'todos'],
             [{tagName: 'input', className: 'new-todo',
               placeholder: 'What needs to be done?', autofocus: true,
               onkeydown: (e, title) =>
               e.key === 'Enter' && (e.target.value = '', fresh((todos, x) =>
                   [m.eq({todos: todos}), x.eq(nil), todos.tailo(x),
                    x.set(list({title: title,
                                done: false, editing: false}))]))}]],
            [{tagName: 'section', className: 'main'},
             [{tagName: 'input', id: 'toggle-all', className: 'toggle-all',
               type: 'checkbox'}],
             [{tagName: 'label', for: 'toggle-all'}, 'Mark all as complete'],
             items_template(m)],
            [{tagName: 'footer', className: 'footer'},
             [{tagName: 'ul', className: 'filters'},
              ['li', [{tagName: 'a', className: 'selected', href: '#/',
                       onclick: m.set({active: true, completed: true})}, 'All']],
              ['li', [{tagName: 'a', href: '#/active',
                       onclick: m.set({active: true, completed: false})},
                      'Active']],
              ['li', [{tagName: 'a', href: '#/completed',
                       onclick: m.set({active: false, completed: true})},
                      'Completed']]],
             [{tagName: 'button', className: 'clear-completed',
               onclick: fresh((todos, item, rest) =>
                   [m.eq({todos: todos}), todos.tailo(item),
                    item.eq(cons({done: true}, rest)), item.set(rest)])},
              'Clear completed']]]; }

function items_template(m) {
    return [{tagName: 'ul', className: 'todo-list'},
            v =>
            fresh((todos, todo, item, rest, title, done, editing, strikethru,
                   active, completed) =>
                [m.eq({todos: todos, active: active, completed: completed}),
                 todos.tailo(item),
                 item.eq(cons(todo, rest)),
                 todo.eq({title: title, done: done, editing: editing}),
                 conde([done.eq(true), completed.eq(true),
                        strikethru.eq('completed')],
                       [done.eq(false), active.eq(true), strikethru.eq('')]),
                 v.eq([{tagName: 'li', className: strikethru},
                       v => [editing.eq(false),
                             v.eq([{tagName: 'div', className: 'view',
                                    ondblclick: e => editing.set(true)},
                                   [{tagName: 'input', id: 'check',
                                     className: 'toggle', type: 'checkbox',
                                     checked: done,
                                     oninput: e => (done.set(e.target.checked))}],
                                   ['label', title],
                                   [{tagName: 'button', className: 'destroy',
                                     onclick: item.set(rest)}]])],
                       v => [editing.eq(true),
                             v.eq([{tagName: 'input', className: 'edit',
                                    value: title,
                                    onkeydown: e => {if (e.key === 'Enter')
                                                     e.target.blur()},
                                    onblur: (e, t) => [editing.set(false),
                                                       title.set(t)]}])]])])]; }
\end{lstlisting}

\begin{figure}[htbp]
\centering
\includegraphics[width=.9\linewidth]{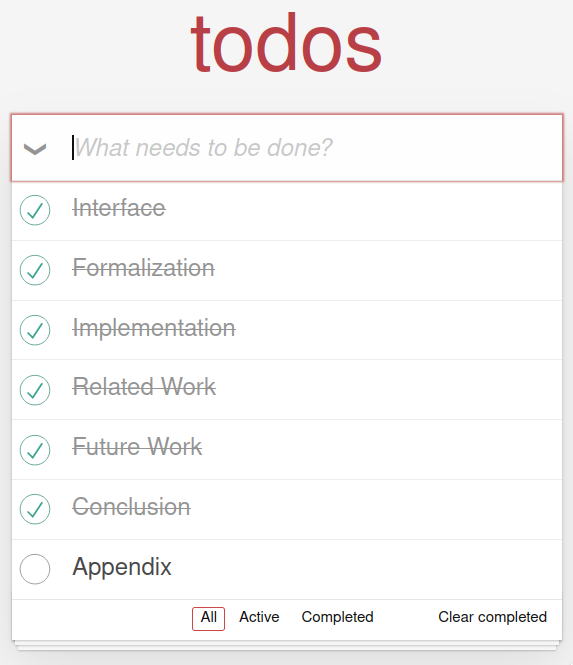}
\caption{\label{fig:orgd36fa6b}
Visual presentation of TodoMVC implementation}
\end{figure}

The final visual result, populated with some example data, can be seen in Figure \ref{fig:orgd36fa6b}.

\end{document}